\documentstyle[12pt,titlepage]{article}

\begin{document}
\input psfig

\title
{A Scaling Theory for
Horizontally Homogeneous, Baroclinically Unstable Flow
on a Beta-Plane}

\author
{Isaac M. Held \\
Geophysical Fluid Dynamics Laboratory/NOAA, Princeton NJ
\and Vitaly D. Larichev\\
Program in Atmospheric and Oceanic Sciences, \\
Princeton University, Princeton NJ}

\maketitle

\begin{abstract}

The scaling argument developed by Larichev and Held (1995) for eddy
amplitudes and fluxes in a horizontally homogeneous, two-layer model on an
$f$-plane is extended to a $\beta$-plane.  In terms of the non-dimensional
number $\xi = U/(\beta \lambda^{2})$, where $\lambda$ is the deformation radius
and \(U\) is the mean thermal wind, the result for the RMS eddy velocity \(V\),
the characteristic wavenumber of the energy-containing eddies and of the
eddy-driven jets $k_{j}$, and the magnitude of the eddy diffusivity for
potential vorticity  \(D\), in the limit $\xi \gg 1$, are as follows:

$ \hspace{1.0in} V/U \approx \xi;
\hspace{0.5in}k_{j}\lambda \approx \xi^{-1} ;
\hspace{0.5in}D/(U\lambda) \approx \xi^{2}$

\noindent
Numerical simulations provide qualitative support for this scaling, but suggest
that it underestimates the sensitivity of these eddy statistics to the value
of $\xi$.
A generalization that is applicable to continuous stratification is
suggested which leads to the estimates:

$\hspace{1.0in} V\approx (\beta T^{2})^{-1};
\hspace{0.5in} k_{j}\approx \beta T ;
\hspace{0.5in}  D\approx (\beta^{2}T^{3})^{-1}$

\noindent
where $T$ is a time-scale determined by the environment; in
particular, it equals
$\lambda U^{-1}$ in the two-layer model and $N(f\partial_{z}U)^{-1}$
in a continuous flow with uniform shear and stratification.  This same
scaling has also been suggested
as relevant to a continuously stratified fluid in the opposite limit,
$\xi \ll 1$.  (Held, 1980).  Therefore, we suggest that it may be of general
relevance in planetary atmospheres and in the oceans.

\end{abstract}

\section{Introduction}

An understanding of the mechanisms which control the amplitudes of baroclinic
eddies and their transport properties is basic to developing
theories for the general circulation of planetary atmospheres and
oceans,
 and for
incorporating the effects of mesoscale eddies in large-scale ocean
models. Different classes of idealized models are needed to improve our
understanding of various aspects of this problem.  We focus on one such
class -- models of horizontally-homogeneous quasi-geostrophic turbulence,
with imposed large-scale potential vorticity gradients.  These models are
useful in bridging the gap between studies of homogeneous two-dimensional
turbulence and studies of baroclinic eddy fluxes in inhomogeneous flows of
physical interest.  Analyses of these baroclinic homogeneous models, and work
on closely related problems, can be found in  Rhines (1977), Salmon
(1978, 1980), Haidvogel and Held (1980), Vallis (1983), Hoyer and
Sadourny (1982), Panetta (1993), and Held and  O'Brien(1992).

Recently, Larichev and Held (1995) (LH hereafter) have considered the
statistically steady state of a two-layer model on an $f$-plane, with an
imposed environmental vertical shear, or interface slope, that results in
equal and opposite potential vorticity gradients in the two layers.
Following the lead of Rhines, Salmon, and Hoyer and Sadourny, they describe
a picture of the barotropic and baroclinic energy cascades that leads to
simple scaling arguments for the energy level and potential vorticity fluxes
in this baroclinically unstable system, given the scale to which the inverse
energy cascade extends.

On a  $\beta$-plane, the barotropic inverse energy cascade is halted when the
characteristic overturning time for the energy-containing eddies becomes
comparable to the inverse of the Rossby wave frequency (Rhines, 1975), at
which scale the energy is channeled into zonal jets.  In a baroclinically
unstable flow, these jets also organize the instability, so that the eddies
form a `storm track' on each jet, as described by Williams (1979) and
Panetta (1993).  By combining the scaling argument of LH with the Rhines
scale at which the inverse cascade halts, we obtain a qualitative theory for
the eddy amplitudes and fluxes on a $\beta$-plane.  We discuss the results
for the two-layer model, describe some numerical solutions that provide
partial support for the theory but also point to some
deficiencies, and then indicate how the argument generalizes to
continuous stratification and arbitrary vertical structure in the mean flow.

\section{The two-layer model}

We consider a two-layer QG model with equal-depth layers when at rest.
The barotropic and baroclinic streamfunctions are denoted by
$\psi \equiv (\psi_1 + \psi_2 )/2$ and $\tau \equiv (\psi_1 - \psi_2 )/2$,
where the subscripts 1 and 2 refer to the upper and lower layers respectively.
The corresponding barotropic and baroclinic velocities are referred to as
$(u_{\psi},v_{\psi})$ and $(u_{\tau},v_{\tau})$
The mean flows in the two-layers $U_i = const$ are specified to provide
a vertical shear
\begin{equation}
U \equiv (U_{1} -U_{2} )/2 > 0
\end{equation}
which results in the mean potential vorticity gradients
$\beta \pm U\lambda^{-2} = U\lambda^{-2}(1 \pm \xi^{-1})$
where $\lambda$ is the internal radius of deformation
and where
\begin{equation}
    \xi \equiv U/(\beta \lambda^2),
\end{equation}
The plus sign refers to the upper layer and the minus sign to the
lower layer,
The lower layer PV gradient is reversed if the shear is large enough that
$\xi > 1$, which is the criterion for instability in an inviscid flow.

The argument in LH for the $f$-plane
can be reformulated as follows.  Start by assuming
that the kinetic energy of the flow is predominately barotropic on scales much
larger than the internal radius of deformation $\lambda$.  On these scales the
barotropic mode evolves as in a two-dimensional flow, cascading energy to
larger scales.  Assume that this inverse energy cascade halts on average at
the wavenumber $k_{0}$.  Given the Kolmogorov energy spectrum of  $k^{-5/3}$,
or any spectrum steeper than $k^{-1}$, the bulk of the energy will be
contained on the scale $k_{0}$.  The {\em baroclinic}
potential vorticity is advected by this nearly barotropic flow on large scales.
Since it does not induce a significant part of the flow by which it is
advected, the baroclinic potential vorticity will behave as a passive tracer,
and will be mixed downgradient by the `turbulent diffusion' engendered by the
barotropic flow.  This mixing will be dominated by the largest scales in the
flow.  Since the magnitude of the ambient potential vorticity gradient is
$U \lambda^{-2}$, the typical size of the eddy baroclinic potential vorticity
will be
\begin{equation}
	q' \approx  k_{0}^{-1}(U \lambda^{-2})
\end{equation}
so that
\begin{equation}
	\overline{v'q'} \approx -D (U \lambda^{-2})
\end{equation}
with
\begin{equation}
	D \approx V k_{0}^{-1}
\end{equation}
where $V$ is the rms {\em barotropic} velocity. Thus, (3) is equivalent
to the statement that the {\em baroclinic} velocities on the energy
containing scale ($ v'_{\tau} \approx k_{0}\lambda^{2}q'$) are of the
order of $U$.

The baroclinic potential vorticity, acting as a passive scalar, will
cascade to smaller scales.  On scales larger than $\lambda$, the variance of
this potential vorticity is dominated by thickness fluctuations, and is
therefore proportional to the potential energy in the baroclinic mode.
The rate of eddy energy production per unit
mass in a horizontally homogeneous two-layer model is
\begin{equation}
  \epsilon = - (U_{1} \overline{v'_{1}q'_{1}} + U_{2}\overline{ v'_{2}
q'_{2}})/2
  = U\overline{v'_{\psi} \tau'}/ \lambda^{2}
  \approx  V k_{0}^{-1} (U / \lambda)^{2}
\end{equation}
In this homogeneous model, the
potential vorticity fluxes in the two layers are equal and opposite and
proportional to the thickness flux.

The baroclinic eddy energy is produced on the largest scale $k_{0}^{-1}$
since it is proportional to the potential vorticity flux.  It then cascades
down to the deformation radius. Energy cannot cascade further since the layers
decouple at smaller scales, leaving 2D flow in which energy cannot
proceed  downscale; instead it is converted to barotropic energy, in
which form it cascades back upscale,  eventually to be dissipated.

The barotropic energy level is determined by the requirement that at
equilibrium the
rate at which energy flows to larger scale in the barotropic mode is equal to
the baroclinic energy production. If the energy cascade extends to the scale
$k_{0}$ and the RMS barotropic velocity is $V$, then by dimensional analysis,
\begin{equation}
	\epsilon \approx V^{3}k_{0}
\end{equation}
an expression familiar from 3D turbulence (e.g., Tennekes (1972)). It is
equivalent to the assumption utilized by LH that an inertial range exists.
{}From (6) and (7), one finds
\begin{equation}
	V \approx (k_{0} \lambda ) ^{-1} U
\end{equation}
The barotropic velocities are larger than the mean shear by the factor
$(k_{0} \lambda) ^{-1}$.  We have assumed that this factor is much greater than
unity so as to have a significant inverse energy cascade. Barotropic
velocities are also larger than baroclinic velocities
on the scale $k_{0}$ by the same factor, so the dominance of advection by
the barotropic flow on large scales is assured.  The corresponding
diffusivity is of the magnitude
\begin{equation}
	D \approx  (k_{0} \lambda) ^{-1} U k_{0}^{-1}
\end{equation}
We also have the result that the eddy available potential energy is
proportional to the barotropic kinetic energy.
\begin{equation}
  \frac{\tau^{2}}{\lambda^{2}} \approx \frac{U^{2}}{(k_{0}\lambda)^2}
  \approx v_{\psi}^2 + u_{\psi}^2
\end{equation}
The argument above is equivalent to that provided by LH.

In the basic $f$-plane numerical experiment analyzed in LH, $k_{0}^{-1}$
is determined by the size of the domain.  On a $\beta$-plane, we expect the
inverse cascade to stop at the Rhines scale, at which rms barotropic velocities
$V$ are comparable to Rossby wave phase speeds
\begin{equation}
	k_{0}^{-2} = V/ \beta
\end{equation}
We ignore the anisotropy of the dispersion relation in this scaling.  By
combining (7) and (11), one obtains the expression for
$k_{0}$ in terms of energy cascade rate discussed by Pelinovskiy (1978) or
Vallis and Maltrud (1991):
$k_{0} \approx (\beta^{3}/\epsilon)^{1/5}$.  (If the
lower boundary is not flat, one should replace $\beta$ by the effective
$\beta$ felt by the barotropic mode in the presence of large-scale bottom
relief,
$H |\nabla (f/H)|$).
As the cascade is halted, the barotropic energy is organized into
jets of this meridional scale.  Panetta (1993) has shown that this relation
accounts quite well for the number of jets produced in a model of homogeneous
geostrophic turbulence in the presence of a mean shear, if V is set equal to
the square root of the total kinetic energy (see his Figure 4).

Our assumption that $(k_{0}\lambda)^{-1}\gg 1$ translates into the statement
that $\xi \gg 1$. We assume that the presence of $\beta$ does not
modify the previous scalings when $\xi$ is large.
Combining  (8) and (11), we have
\begin{equation}
	 V/U \approx \xi
\end{equation}
and
\begin{equation}
	k_{0} \lambda \approx \xi^{-1}
\end{equation}
with the consequence that
\begin{equation}
	D \approx U \lambda \xi^{2}
\end{equation}
Since D is proportional to $U^{3}$, the potential vorticity flux is
proportional to $U^{4}$ while the energy generation is proportional to $U^{5}$.

\section{Some numerical results}

These scaling relations can be tested against numerical experiments
such as those described by Haidvogel and Held (1980) and Panetta (1992),
for 2-layer, QG homogeneous turbulence on a $\beta$-plane with imposed mean
vertical shear.  Figure 18 in Panetta (1993) shows that the scale with the
maximum
energy and the scale of maximum
energy generation are both approximately linear in $\xi$, consistent with (13).
(In the notation of Panetta, $\xi = (2 \beta)^{-1}$.)
The predicted eddy amplitudes and fluxes are not that well verified, however.
The eddy energy should be proportional to $\xi^{2}$.  The results tabulated by
Panetta suggest that the energies and fluxes are even more sensitive
to $\xi$. To confirm this result, we have obtained numerical
solutions to the same model as considered by Haidvogel and Held (1980) and
Panetta(1993), but with 256x256 resolution that allows more scale separation
between the radius of deformation and the energy containing eddy scale.

The numerical model is identical to that used in the $f$-plane simulations
of LH.  We choose a value of the lower layer Ekman damping,
$\kappa =  0.16 (U/\lambda )$ that is identical to that in LH, and set the
deformation radius so that $2\pi \lambda = L/50$,
where L is the size of the square doubly-periodic domain.  Small scale
damping is modeled with a $\nu \nabla^{8}$ diffusion operator, with
$\nu  = 0.08 (U \lambda^{7})$.  Statistically steady states have been obtained
for the values of $\xi$ shown in Table 1. Instability actually persists for
values of $\xi$ slightly less than 1, due to dissipative destabilization, but
these are not displayed in the table.

Figure 1 shows the horizontal spectra of the barotropic component of the
meridional velocity and of the
baroclinic eddy energy generation (or equivalently, of the potential vorticity
flux in either layer, or of the buoyancy flux) for each of the statistically
steady states obtained. The spectra are plotted as a function of total
horizontal wavenumber, after averaging over angle in k-space.
{\em The energy generation clearly moves to larger scales when the
energy is allowed to cascade
to larger scales.} Also shown by arrows is the Rhines scale (11), computed
using the rms barotropic meridional velocity for $V$, for each value of $\xi$.
This prediction of the scale is seen to be in reasonable agreement with
the scale of the maxima in energy and energy generation, for large values
of $\xi$ down to $\xi \approx 2$.

As Fig.2 shows, the energies and the energy generation increase more
rapidly with increasing $\xi$ than predicted by our scaling arguments.
Instead of the expected $\xi^2$ power law, we obtain a dependence
that is between $\xi^3$ and $\xi^4$.
This result confirms the impression obtained from Panetta's results.
\begin{table}[t]
\begin{tabular}{|c||c|c|c|c|c|}
\hline

$\xi$
& $\overline{v_{\psi}^{\prime \ 2}}/U^{2}$
& $\overline{v_{\psi}^{\prime \ 2}}/\overline{u_{\psi} ^{\prime \ 2}}$
& $\epsilon/(U^{3}\lambda^{-1})$
& $\epsilon_{diss}/\epsilon$
& $(k_{0} \lambda)^{-1}$  \\
\hline
8     & 488    & 0.92   &  104     &  0.74   &  13.3  \\
6     & 175    & 0.88   &  38.4    &  0.70   &  8.91  \\
4     & 38.6   & 0.76   &  8.70    &  0.69   &  5.00  \\
3     & 12.7   & 0.62   &  3.01    &  0.69   &  3.27  \\
2     & 2.85   & 0.48   &  0.717   &  0.69   &  1.84  \\
10/7  & 0.77   & 0.40   &  0.211   &  0.64   &  1.12  \\
10/9  & 0.28   & 0.33   &  0.085   &  0.58   &  0.77  \\
\hline
\end{tabular}
\caption{Statistics from the numerical model as a function of $\xi$:
$\epsilon_{diss}$ is the dissipation of energy in the barotropic mode by
Ekman friction; $k_{0}$ is computed from (11) using
the rms meridional velocity of the barotropic mode for V.}

\end{table}
We cannot rule out the possibility that even the largest values of $\xi$
are not yet large enough to attain the asymptotic regime predicted, but
this appears
to be unlikely since there is no trend in the results that suggests that the
$\xi^{2}$ behavior is being approached.  (The comparison with theory is also
made more complicated by the fact that the size of the domain is beginning to
make itself felt at the largest value of $\xi$ used.)  This
discrepancy is also seen in the $f$-plane results in LH, in which the
scale of the energy generation was arbitrarily constrained.

To examine the sources of this discrepancy, we first plot in figure 3 the
energy generation $\epsilon$ normalized by
\begin{equation}
  \frac{U}{2 \lambda^{2}} (\overline{v_{\psi}^{\prime \ 2}})^{1/2}
   (\overline{ \tau^{\prime \  2}})^{1/2}
\end{equation}
This quantity equals unity
if the meridional velocity and thickness are perfectly correlated.  The
values obtained are close to 0.2 and hardly change except for the
largest value of $\xi$ examined.  Therefore, the neglect of changing
correlations in the estimate (6) does not appear to be a significant source of
error.

Combining (6) with the estimate (11) for the mixing length, we have
\begin{equation}
\frac{\epsilon}{U^{3}\lambda^{-1}} \approx  (\frac{V}{U})(k_{0}\lambda)^{-1}
\approx (\frac{V}{U})^{3/2} \xi^{1/2} \equiv
\frac{\epsilon_1}{U^{3}\lambda^{-1}}
\end{equation}
Alternatively, combining (7) with (11) yields
\begin{equation}
\frac{\epsilon}{U^{3}\lambda^{-1}} \approx (\frac{V}{U})^{3}(k_{0}\lambda)
\approx (\frac{V}{U})^{5/2}\xi^{-1/2} \equiv
\frac{\epsilon_2}{U^{3}\lambda^{-1}}
\end{equation}
These two expressions for $\epsilon$ can be combined to yield (12)-(14).
We plot $\epsilon/\epsilon_1$ and $\epsilon/\epsilon_2$ in Figure 4a as a
function of $\xi$.  Once again V has been set equal to the rms barotropic
meridional velocity.  Considering that $\epsilon$ varies by more than 3 orders
of magnitude (see table), the constancy of these ratios is encouraging.
However, it is
apparent that the RHS of (16) underestimates the dependence of the energy
production on $\xi$, indicating that the effective mixing length increases
somewhat more rapidly with $\xi$ than does the Rhines scale.  In contrast,
(17) overestimates the dependence of $\epsilon$ on $\xi$, indicating that the
inverse energy cascade is not as efficient as indicated by (7) if one uses the
Rhines scale for $k_{0}$.

In Fig. 4a we have also compared the estimate (17)
to the simulated energy dissipation by
Ekman damping in the barotropic mode rather than to the eddy energy
production.  These are implicitly assumed to be equal, or at least
proportional, in our scaling
arguments, but they differ in the numerical model
because of the energy lost due to the subgrid
diffusivity and the Ekman damping of
baroclinic kinetic energy (see Table 1).
The dissipation in the barotropic mode is
a better estimate of the rate at which energy is cascading to larger scales,
which is the more relevant quantity for comparison with (17).  The curve
becomes
a bit flatter at smaller values of $\xi$, but most of the discrepancy
remains.  Although we have compromised our resolution at small scales in order
to provide room for a substantial inverse cascade, resulting in non-trivial
energy losses due to the subgrid scale diffusivity, we do not think that
a model with higher resolution and weaker diffusion would produce qualitatively
different results.

If we use the total (zonal plus meridional) rms eddy velocity for $V$ in (7),
while retaining the use of the meridional velocity in (11), the plot
analogous to
that in Figure 4a (dashed line) shows a much improved fit, as shown in
Figure 4b.  This
choice is admittedly somewhat arbitrary in the absence of a theory for
the anisotropy of the energy spectrum, but it does indicate that the
anisotropy of the eddies (see the ratio of rms eddy $u_{\psi}$ and $v_{\psi}$
in the table),
which we have ignored in our simple scaling, likely plays a role in the
discrepancy in Fig. 4a.

Because (16) underestimates the energy production and (17) overestimates it,
these errors add to produce the larger errors in the dependence of the energy
level and fluxes on $\xi$.  For example, if one arbitrarily replaces (16) and
(17) with
\begin{equation}
\frac{\epsilon}{U^{3}\lambda^{-1}} \approx (\frac{V}{U})^{3/2+\alpha} \xi^{1/2}
\end{equation}
\begin{equation}
\frac{\epsilon}{U^{3}\lambda^{-1}} \approx (\frac{V}{U})^{5/2-\gamma}\xi^{-1/2}
\end{equation}
then one finds
\begin{equation}
(V/U)^{2} \approx \xi^{\delta}; \delta \equiv 2/(1-\alpha -\gamma)
\end{equation}
Choosing $\alpha = \gamma  = 0.20$ or $0.25$, as Fig.4a roughly suggests,
one obtains $\delta = 10/3 $ or $4$,
consistent with the variation of energy with $\xi$ in Figure 2.
Errors that
appear to be small in the individual approximations combine to create a
large discrepancy in the final dependence of energy on $\xi$.  The underlying
reason for this sensitivity is the strong positive feedback inherent in the
system:
as the energy level increases, the length scale of the energy containing eddies
increases, which increases the mixing length and V, increasing the energy
generation.

Although our qualitative arguments clearly require some modification
in order to fit numerical results, we feel that they are a useful starting
point, and therefore have examined how they might be generalized to
horizontally homogeneous flows with arbitrary stratification and mean vertical
shears.

\section{Continuous stratification}

Consider a Boussinesq QG flow with stratification $N^{2}(z)$
and mean flow
$U(z)$, confined between flat horizontal boundaries at $z=0,H$.  The energy
production per unit mass $\epsilon$  can be written in the form
\begin{equation}
\epsilon =  H^{-1}\int_{0}^{H}
 \frac{f\overline{v'b'}}{N^{2}} \frac {\partial U}{\partial z}dz
\end{equation}
Here $\overline{v'b'}$ is the meridional buoyancy flux, related to the
potential vorticity flux by
\begin{equation}
	\overline{v'q'} = f\partial_{z}(N^{-2} \overline{v'b'})
\end{equation}
Consistent with our assumption of horizontal homogeneity, the relative
vorticity flux, which equals the convergence of the eddy momentum fluxes,
is ignored.  The mean wind shear is related to the mean buoyancy gradient by
\begin{equation}
	f \partial_z U = - \partial_y B
\end{equation}

We assume again that the inverse cascade is substantial enough that
the flow advecting the potential vorticity can be taken
as barotropic.   The downgradient fluxes can then be set equal to a
diffusivity $D$ times the negative of the mean gradient, with the diffusivity
independent of $z$. In this case of a vertically uniform diffusivity, the
same diffusivity can be applied to buoyancy and to potential vorticity.
The result is
\begin{equation}
          \epsilon = D T^{-2}
\end{equation}
where
\begin{equation}
	T^{-2} \equiv H^{-1} \int_{0}^{H}
 \frac{f^{2}}{N^{2}} (\frac {\partial U}{\partial z})^{2} dz
=H^{-1}f^{2} \int_{0}^{H} Ri^{-1} dz
\end{equation}
For the two layer model of Section 2, $T = \lambda /U$,where
$U = (U_{1}-U_{2})/2$, from which we can rederive the 2-layer scaling.
For a continuous flow with $N$ and $\partial_z U$ constants, we have
$T = N(f\partial_zU)^{-1} = f^{-1} Ri^{1/2}$.
In both cases, $T$  is proportional to the e-folding time for baroclinic waves
in the limit that the effect of $\beta$ is small.
	We again assume that the diffusivity $D$ is related to the energy
containing scale $k_{0}$ and the barotropic velocity scale $V$, as in (5), and
that
the energy production must balance the upscale transport of energy by the
barotropic flow (7), so that
\begin{equation}
	\epsilon = Vk_{0}^{-1}T^{-2}  = V^{3}k_{0},
\end{equation}
or simply,
\begin{equation}
	T = 1/(k_{0}V)
\end{equation}
Thus, in addition to being the time scale associated with the linear dynamics
of deformation scale eddies, T is also the time scale of the energy containing
eddies on much larger scales.  Combined with the estimate of $k_{0}$
as the Rhines scale (11), we have
\begin{equation}
	k_{0} = \beta T,
\end{equation}
\begin{equation}
	V = 1/(\beta T^{2}),
\end{equation}
\begin{equation}
	D = 1/(\beta^{2}T^{3})
\end{equation}

\section{Relationship to more weakly unstable flows}

The preceding analysis is intended for the asymptotic limit of small
$\beta$, in which there is an inverse energy cascade over a substantial range
of
scales.  It is in this limit that the flow responsible for diffusing the
potential vorticity is primarily barotropic, so that we can remove the
diffusivity $D$ from the integral in the expression for the energy generation
(21). In the two-layer model, these scaling relations cannot remain
qualitatively valid for all $\xi$, since the flow is stable once $\xi$ drops
below a critical value (although it is impressive how well the two-layer
results in Figure 2 are approximated by a power law even as $\xi$ approaches
unity).  However, these expressions may retain their relevance
over the full range of $\xi$ in certain continuous flows.  Held (1978) suggests
scaling arguments for Charney-like continuous models in the limit that
$h/H \ll 1$,
where $H$ is the depth of the fluid, or the scale height in the non-Boussinesq
case, while
\begin{equation}
	h \equiv \frac{f^{2}\partial_{z}U}{\beta N^{2}}
\end{equation}
In this limit the vertical scale of the most unstable waves in Charney's model
is proportional to $h$, and their horizontal scale to
\begin{equation}
	Nh/f = \frac{f\partial_zU}{\beta N} \equiv 1/(\beta T)
\end{equation}
where we have used the expression (25) for $T$.  This is identical to the
scaling in (28).  If these scales dominate the statistically steady state,
surface buoyancy perturbations will be of magnitude
$b' \approx (Nh/f) \partial_{y} B$,
or, using equipartition of eddy kinetic and available potential energies,
\begin{equation}
V \approx b'/N \approx h\partial_{z}U \approx \beta^{-1}(f\partial_{z}U/N)^{2}
\end{equation}
which is identical to (29).  The result is again a diffusivity proportional to
the cube of the vertical shear.

In Held (1978), this scaling was assumed to break down as $h/H$ becomes
comparable to  unity.  Following Stone (1972), it was assumed that the
dominant horizontal scale relevant for the eddy transports would then be
$NH/f = \lambda$ and the characteristic velocity $H\partial_{z}U$,
leading to a diffusivity that
is proportional to the first power of the vertical shear.  The argument of LH
suggests instead that once $h/H (= \xi )$ rises above unity, the inverse energy
cascade sets in and the appropriate eddy scale and mixing length continue to
increase in proportion to $\xi$.  With this picture in mind, one can
imagine that the scalings (28-30) are relevant for the full range of $\xi$.

In the weakly unstable case one cannot assume
that the diffusivity is independent of height.  In fact, if $h/H \ll 1$
the mixing should be confined to a depth proportional to $h$ (Held, 1978) in a
Charney-like environment.  More generally, a theory for the vertical structure
of the diffusivity is required as one moves away from the strongly unstable
limit.

\section{Final remarks}

The qualitative theory presented for the eddy amplitudes and scales in
horizontally homogeneous baroclinically unstable flows  can be most simply
described by the statement that the environmental shear and stratification
determine a time scale $T$, which can be combined with $\beta$
to determine a length $(\beta T)^{-1}$,
a velocity $(\beta T^{2})^{-1}$,
and a diffusivity $(\beta^{2}T^{3})^{-1}$ in only one way.
The remainder of the argument provides the rationale for thinking that the
shear and stratification provides a time scale only, in the limit in which the
flow is energetic enough that there exists a substantial inverse energy
cascade.  The result coincides in form to that proposed by Held
(1978) for a Charney-like environment, in the opposite limit in which the
eddies are very weak.  This leads us to suggest that these scalings may in
fact be useful for a wide range of energy levels in certain flows,
although we do not understand if there are fundamental reasons why
this should be so.

In practice, one must keep in mind that there may be insufficient room for
$\beta$ to halt the cascade, in which case the
eddy scale is determined by the domain size, or the size of the baroclinic
region.  This leads to the diffusivity proposed by Green (1970), as described
by LH.

Several limitations to these arguments are evident.  Even in the limit
of very strong supercriticality in the two layer model, there is only
partial agreement with numerical simulations.  Our scaling arguments suggest
that an accurate theory for eddy amplitudes and fluxes will be difficult to
obtain, because of the delicate balance resulting from a 0 feedback
in which potential vorticity fluxes increase as eddy velocity and
length scales increase, leading to higher energy levels and a
stronger inverse energy cascade, leading, in turn, to an increase in
eddy velocity and length scales.

Even if the limitations of the asymptotic theory are ignored, one still needs
a theory for the vertical structure of the diffusivity, or of the eddy PV
fluxes, when the inverse energy cascade is not so strong as to generate a
nearly barotropic eddy velocity field.  These arguments also do not tell us
how to treat the dependence of eddy statistics on the strength of the surface
drag.  Also, the bottom topography so far ignored in the analysis may strongly
effect the cascades, rendering the flows more baroclinic (Cox, 1985;
Treguier and Hua, 1988).

The final and most fundamental limitation relates to the assumption of
horizontal homogeneity.  For example, if the barotropic decay is due to
momentum fluxes in the presence of a mean horizontal shear, and if this
horizontal shear is imposed by
other constraints, rather than being internally generated by the eddies
themselves, then  the dependence of the length and velocity scales on
environmental parameters would clearly be changed.  However, the recent study
of Pavan and Held (1995) suggests that diffusivities obtained from a
homogeneous model can be surprisingly useful for jet-like flows with
substantial horizontal shears.

Despite these limitations and misgivings, we believe that
these scaling arguments provide a useful background for developing closure
schemes for simple diffusive atmospheric models as well as for non-eddy
resolving ocean models.

\vspace{5cm}
{\large \bf \noindent Acknowledgments}

\vspace{1cm}
The authors would like to thank Isidoro Orlanski and Valentina Pavan for
helpful discussions on this and related topics.  VDL was funded under a
grant/cooperative agreement with the National Oceanic and Atmospheric
Administration (NA26RG0102-01).  The views expressed herein are those of
the authors and do not necessarily reflect the views of NOAA or any of its
subagencies.

\newpage
\centerline{\large \bf References}
\vspace{1cm}

\begin{description}
\item[]
Cox, M.D., 1985: An eddy-resolving numerical model of the ventilated
thermocline.
{\em J. Phys. Oceanog.}, {\bf 15}, 1312-1324.
\item[]
Green, J.S.A., 1970: Transfer properties of the large-scale eddies and
the general circulation of the atmosphere.
{\em Quart. J. Roy. Meteor. Soc.}, {\bf 96}, 157-185.
\item[]
Haidvogel, D.B., and I.M. Held, 1980: Homogeneous quasi-geostrophic turbulence
driven by a uniform temperature gradient.
{\em J. Atmos. Sci.}, {\bf 37}, 2644-2660.
\item[]
Held, I.M., 1978:  The vertical scale of an unstable baroclinic wave and its
importance for eddy heat flux parameterizations.
{\em J. Atmos. Sci.}, {\bf 35}, 572-576.
\item[]
----------, and E. O'Brien, 1992: Quasigeostrophic turbulence in a three-layer
model: effects of vertical structure in the mean shear.
{\em J. Atmos. Sci.}, {\bf 49}, 1861-1870.
\item[]
Hoyer, J.-M., and R. Sadourny, 1982: Closure modeling of fully developed
baroclinic instability.
{\em J. Atmos. Sci.}, {\bf 39}, 707-721.
\item[]
Larichev, V.D., and I.M.Held, 1995: Eddy amplitudes and fluxes in a
homogeneous model of fully developed baroclinic instability.
Submitted to {\em J.Phys. Oceanog.}
\item[]
Maltrud, M.E., and G.K.Vallis, 1991: Energy spectra and coherent structures in
forced two-dimensional and beta-plane turbulence.
{\em J.Fluid. Mech.}, {\bf 228}, 321-342.
\item[]
Panetta, R.L., 1993: Zonal jets in wide baroclinically unstable regions:
persistence and scale selection.
{\em J. Atmos. Sci.}, {\bf 50}, 2073-2106.
\item[]
Pavan, V., and I.M.Held, 1995: The diffusive approximation for eddy
fluxes in baroclinically unstable jets.
Submitted to {\em J. Atmos. Sci.}
\item[]
Pelinovskiy, Ye.N., 1978: Wave turbulence on a beta-plane.
{\em Oceanology}, {\bf 18}, 126-128
\item[]
Rhines, P.B., 1977: The dynamics of unsteady currents. {\em The Sea, Vol. 6},
E.A. Goldberg, I.N. McCane, J.J. O'Brien and J.H.Steele, Eds., Wiley, 189-318.
\item[]
Rhines, P.B., 1975: Waves and turbulence on a beta-plane.
{\em J.Fluid.Mech.}, {\bf 69}, 417-443.
\item[]
Salmon, R., 1978: Two-layer quasi-geostrophic turbulence in a simple special
case.
{\em Geophys.Astrophys. Fluid Dynamics}, {\bf 10}, 25-52.
\item[]
----------, 1980: Baroclinic instability and geostrophic turbulence.
{\em Geophys.Astrophys. Fluid Dynamics}, {\bf 15}, 167-211.
\item[]
Stone, P.H., 1972:  A simplified radiative-dynamical model for the static
stability of rotating atmospheres.
{\em J. Atmos. Sci.}, {\bf 29}, 405-418.
\item[]
Tennekes, H., 1972: {\em A First Course in Turbulence},
MIT Press, Cambridge, MA,
\item[]
Treguier, A.M. and B.L.Hua, 1988: Influence of bottom topography on
stratified quasi-geostrophic turbulence in the ocean.
{\em Geophys. Astrophys. Fluid Dynamics}, {\bf 43}, 265-305.
\item[]
Vallis, G.K., 1983: On the predictability of quasi-geostrophic flow: the
effects of beta and baroclinicity.
{\em J. Atmos. Sci.}, {\bf 40}, 10-27.
\item[]
Williams, G. P., 1979:  Planetary circulations 2:  The Jovian quasi-geostrophic
regime.
{\em J. Atmos. Sci.}, {\bf 36}, 932-968.

\end{description}

\newpage
\centerline{\large \bf Figure Captions}
\vspace{1cm}

\begin{description}
\item[]
Fig. 1. The horizontal spectra of (a) the barotropic meridional
velocity and (b) the baroclinic eddy energy production rate, for different
$\xi$ at equilibrium.  The arrows indicate the Rhines scale computed from
(11) using the rms barotropic meridional velocity for V.
\item[]
Fig. 2. A log-log plot of $\overline{v_{\psi}^{\prime \ 2}}$,
twice the meridional barotropic
energy (solid line),  and $\epsilon$, the baroclinic
eddy energy production rate (dashed line), as functions of the
supercriticality $\xi$. Also shown are slopes consistent with $\xi ^2$ and
$\xi ^4$ dependencies.
\item[]
Fig. 3. The baroclinic eddy energy production rate, $\epsilon$
as a function of supercriticality $\xi$, normalized by (15).
\item[]
Fig. 4. Ratios of the theoretical estimates (16),(17) to the
numerical simulations, as a function of the supercriticality $\xi$: (a)
$\epsilon/\epsilon_1$ (solid line),
$\epsilon/\epsilon_2$ (dashed line).
and $\epsilon_{diss}/\epsilon_2$ (dotted line).  See (16) and (17) for
definitions of $\epsilon_1$ and $\epsilon_2$.  $\epsilon_{diss}$ is the
energy dissipation by Ekman damping in the barotropic mode;
(b)  $\epsilon/\epsilon_2$ modified by using the total rms eddy barotropic
velocity in (17) (rather than the meridional velocity only, as for the
dashed line in (a)).
\end{description}

\newpage
\centerline{\psfig{figure=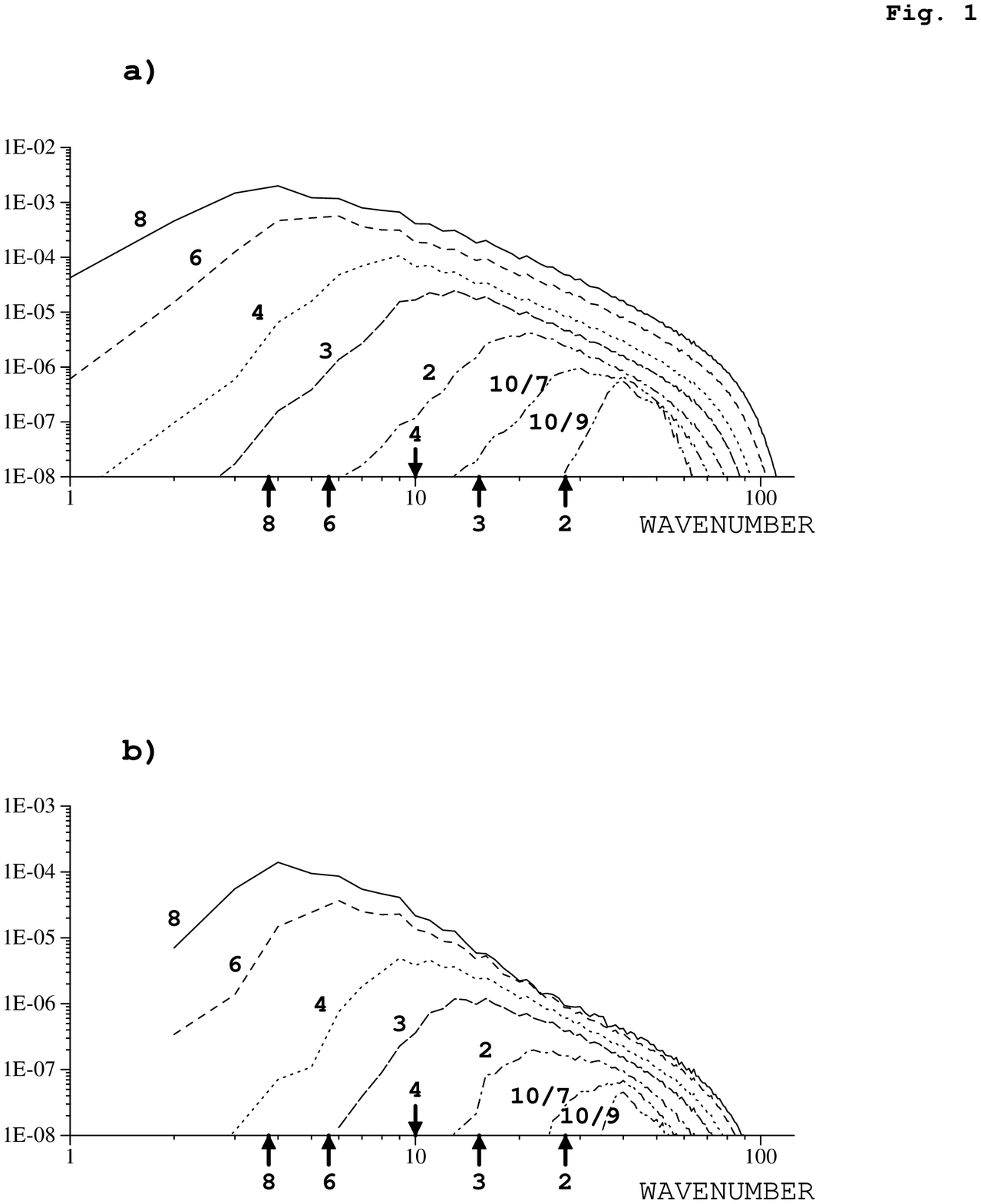,height=7.0in}}
\begin{figure}[htbp] \caption{
  \baselineskip 3ex
The horizontal spectra of (a) the barotropic meridional
velocity and (b) the baroclinic eddy energy production rate, for different
$\xi$ at equilibrium.  The arrows indicate the Rhines scale computed from
(11) using the rms barotropic meridional velocity for V.}
\label{fig:spectra}
\end{figure}

\newpage
\centerline{\psfig{figure=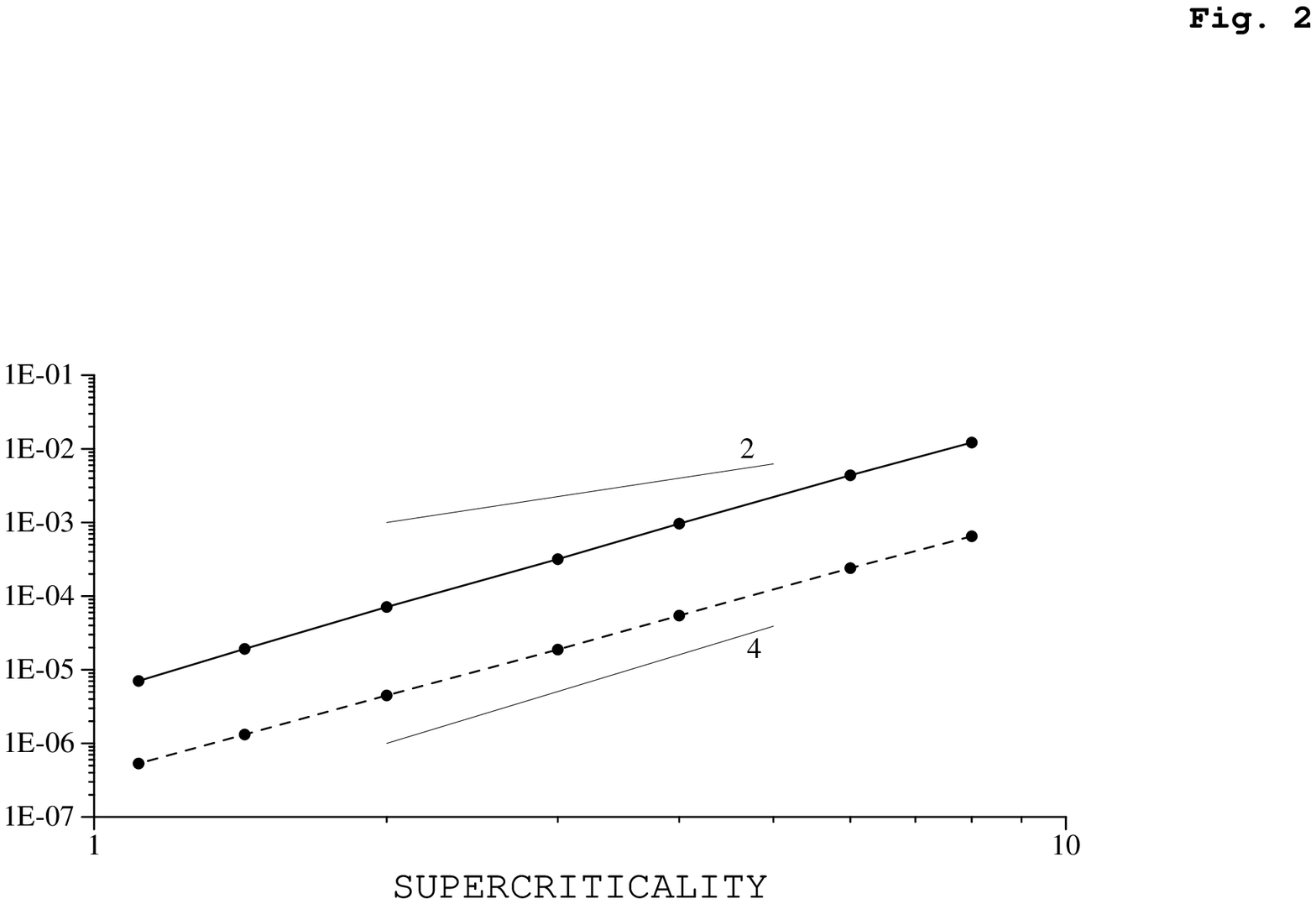,height=5.0in}}
\begin{figure}[htbp] \caption{
  \baselineskip 3ex
A log-log plot of $\overline{v_{\psi}^{\prime \ 2}}$,
twice the meridional barotropic
energy (solid line),  and $\epsilon$, the baroclinic
eddy energy production rate (dashed line), as functions of the
supercriticality $\xi$. Also shown are slopes consistent with $\xi ^2$ and
$\xi ^4$ dependencies.
}
\label{fig:slopes}
\end{figure}

\centerline{\psfig{figure=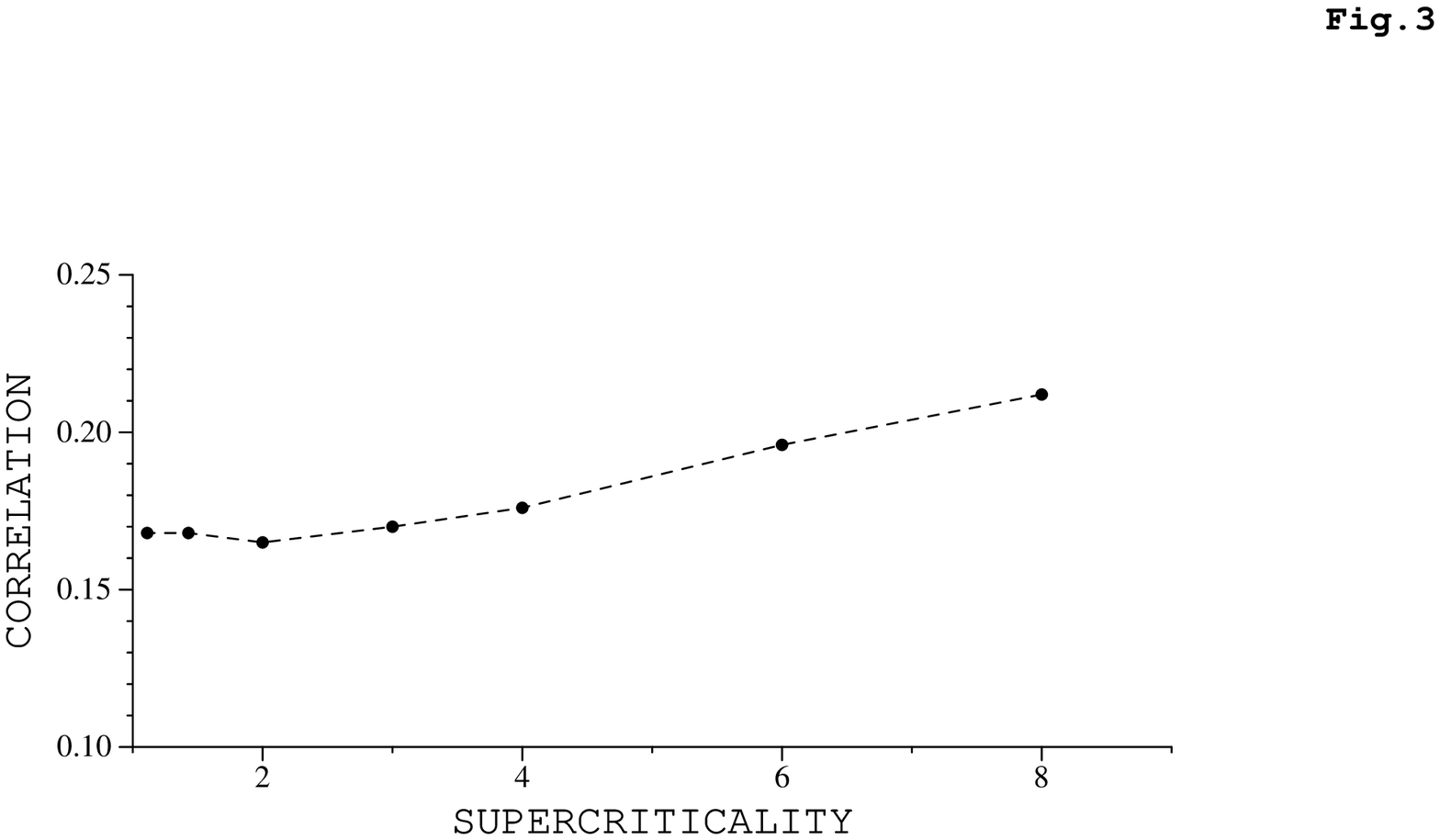,height=5.0in}}
\begin{figure}[htbp]
\caption{
  \baselineskip 3ex
The baroclinic eddy energy production rate, $\epsilon$
as a function of supercriticality $\xi$, normalized by (15).
}
\label{fig:correlation}
\end{figure}

\newpage
\centerline{\psfig{figure=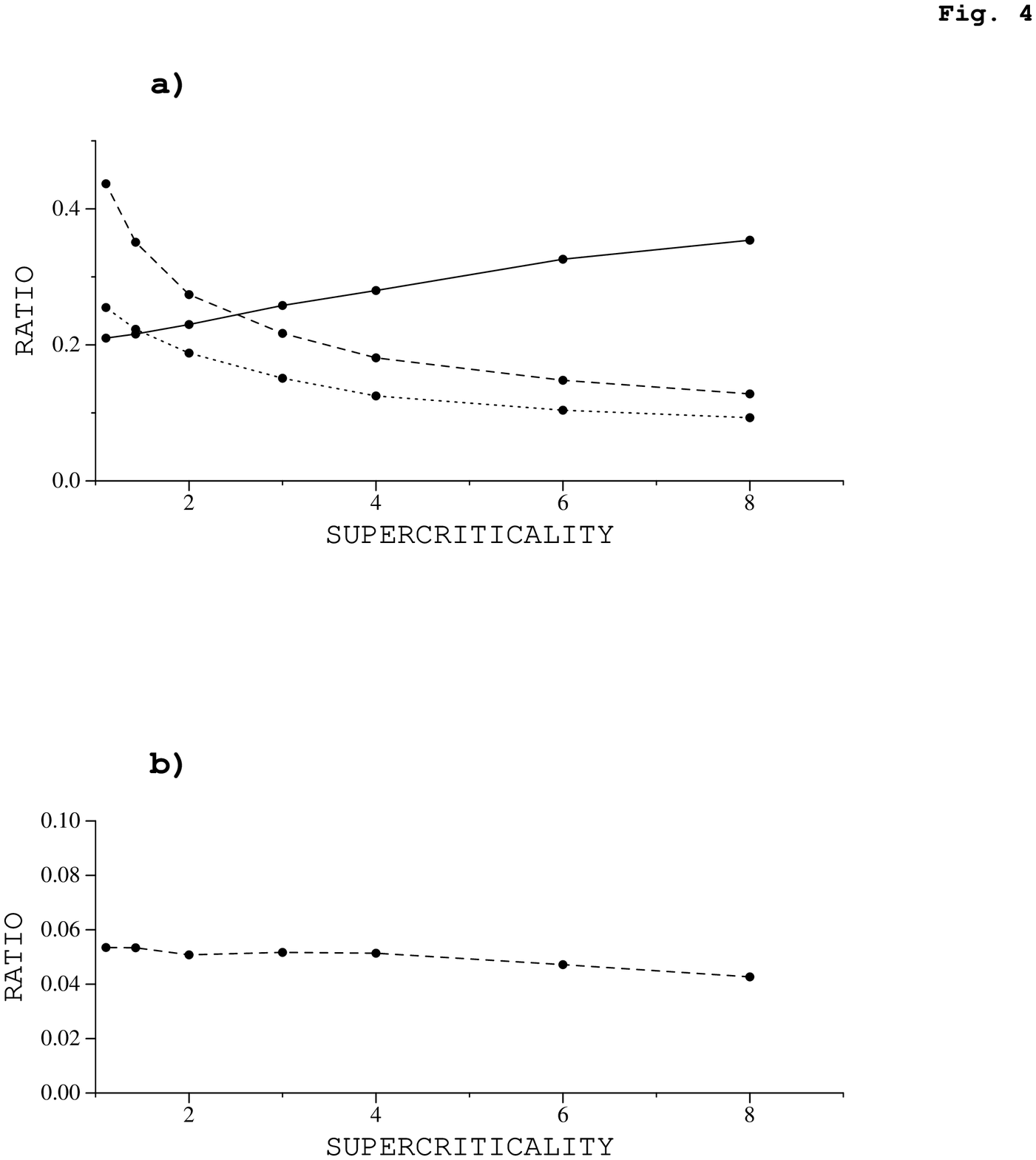,height=7.0in}}
\begin{figure}[htbp] \caption{
  \baselineskip 3ex
Ratios of the theoretical estimates (16),(17) to the
numerical simulations, as a function of the supercriticality $\xi$: (a)
$\epsilon/\epsilon_1$ (solid line),
$\epsilon/\epsilon_2$ (dashed line).
and $\epsilon_{diss}/\epsilon_2$ (dotted line).  See (16) and (17) for
definitions of $\epsilon_1$ and $\epsilon_2$.  $\epsilon_{diss}$ is the
energy dissipation by Ekman damping in the barotropic mode;
(b)  $\epsilon/\epsilon_2$ modified by using the total rms eddy barotropic
velocity in (17) (rather than the meridional velocity only, as for the
dashed line in (a)).
}
\label{fig:decomposition}
\end{figure}

\end{document}